# Non-degenerate 2-photon excitation in scattering medium for fluorescence microscopy

Mu-Han Yang,[1]* Maxim Abashin,[1] Payam A Saisan,[2] Peifang Tian,[2,4] Christopher G.L. Ferri[2], AnnaDevor,[2,3,5] and Yeshaiahu Fainman[1]

[1]*Department of Electrical and Computer Engineering, University of California, San Diego, CA 92093, USA*
[2]*Department of Neurosciences, University of California, San Diego, CA 92093, USA*
[3]*Department of Radiology, University of California, San Diego, CA 92093, USA*
[4]*Department of Physics, John Carroll University, University Heights, OH 44118*
[5]*Martinos Center for Biomedical Imaging, Massachusetts General Hospital, Charlestown, MA02129*
[*]*muy003@eng.ucsd.edu*

**Abstract:** Non-degenerate 2-photon excitation (ND-2PE) of a fluorophore with two laser beams of different photon energies offers an independent degree of freedom in tuning of the photon flux for each beam. This feature takes advantage of the infrared wavelengths used in 3-photon microscopy to achieve an increased penetration depth, while preserving a relatively high degenerate 2-photon excitation (D-2PE) cross section, exceeding that achievable with 3-photon excitation. Here, using spatially and temporally aligned Ti:Sapphire laser and optical parametric oscillator beams operating at near infrared (NIR) and short-wavelength infrared (SWIR) optical frequencies, respectively, we provide a practical demonstration that the emission intensity of a fluorophore excited in the non-degenerate regime in a scattering medium is more efficient than the commonly used D-2PE.

**References and links**

1. K. Svoboda and R. Yasuda, "Principles of two-photon excitation microscopy and its applications to neuroscience," Neuron **50**, 823–839 (2006).
2. J. N. Kerr and W. Denk, "Imaging in vivo: watching the brain in action," Nat. Rev. Neurosci. **9**, 195–205 (2008).
3. W. Denk and K. Svoboda, "Photon upmanship: why multiphoton imaging is more than a gimmick," Neuron **18**, 351–357 (1997).
4. F. Helmchen and W. Denk, "Deep tissue two-photon microscopy," Nat. Methods **2**, 932–940 (2005).
5. T. R. Insel, S. C. Landis, F. S. Collins, and others, "The NIH brain initiative," Science **340**, 687–688 (2013).
6. P. Theer, M. T. Hasan, and W. Denk, "Two-photon imaging to a depth of 1000 μm in living brains by use of a Ti: Al 2 O 3 regenerative amplifier," Opt. Lett. **28**, 1022–1024 (2003).
7. K. A. Kasischke, E. M. Lambert, B. Panepento, A. Sun, H. A. Gelbard, R. W. Burgess, T. H. Foster, and M. Nedergaard, "Two-photon NADH imaging exposes boundaries of oxygen diffusion in cortical vascular supply regions," J. Cereb. Blood Flow Metab. **31**, 68–81 (2011).
8. T. Takano, G.-F. Tian, W. Peng, N. Lou, D. Lovatt, A. J. Hansen, K. A. Kasischke, and M. Nedergaard, "Cortical spreading depression causes and coincides with tissue hypoxia," Nat. Neurosci. **10**, 754–762 (2007).
9. K. Nizar, H. Uhlirova, P. Tian, P. A. Saisan, Q. Cheng, L. Reznichenko, K. L. Weldy, T. C. Steed, V. B. Sridhar, C. L. MacDonald, and others, "In vivo stimulus-induced vasodilation occurs without IP3 receptor activation and may precede astrocytic calcium increase," J. Neurosci. **33**, 8411–8422 (2013).
10. D. Kobat, M. E. Durst, N. Nishimura, A. W. Wong, C. B. Schaffer, and C. Xu, "Deep tissue multiphoton microscopy using longer wavelength excitation," Opt. Express **17**, 13354–13364 (2009).
11. D. Kobat, N. G. Horton, and C. Xu, "In vivo two-photon microscopy to 1.6-mm depth in mouse cortex," J. Biomed. Opt. **16**, 106014–106014 (2011).
12. R. Kawakami, K. Sawada, A. Sato, T. Hibi, Y. Kozawa, S. Sato, H. Yokoyama, and T. Nemoto, "Visualizing hippocampal neurons with in vivo two-photon microscopy using a 1030 nm picosecond pulse laser," Sci. Rep. **3**, (2013).
13. P. Theer and W. Denk, "On the fundamental imaging-depth limit in two-photon microscopy," JOSA A **23**, 3139–3149 (2006).
14. N. G. Horton, K. Wang, D. Kobat, C. G. Clark, F. W. Wise, C. B. Schaffer, and C. Xu, "In vivo three-photon microscopy of subcortical structures within an intact mouse brain," Nat. Photonics **7**, 205–209 (2013).
15. C. Xu, R. Williams, W. Zipfel, and W. W. Webb, "Multiphoton excitation cross-sections of molecular fluorophores," Bioimaging **4**, 198–207 (1996).

16. C. Xu, W. Zipfel, J. B. Shear, R. M. Williams, and W. W. Webb, "Multiphoton fluorescence excitation: new spectral windows for biological nonlinear microscopy," Proc. Natl. Acad. Sci. **93**, 10763–10768 (1996).
17. R. P. Barretto, T. H. Ko, J. C. Jung, T. J. Wang, G. Capps, A. C. Waters, Y. Ziv, A. Attardo, L. Recht, and M. J. Schnitzer, "Time-lapse imaging of disease progression in deep brain areas using fluorescence microendoscopy," Nat. Med. **17**, 223–228 (2011).
18. A. Mizrahi, J. C. Crowley, E. Shtoyerman, and L. C. Katz, "High-resolution in vivo imaging of hippocampal dendrites and spines," J. Neurosci. **24**, 3147–3151 (2004).
19. L. V. Wang and S. Hu, "Photoacoustic tomography: in vivo imaging from organelles to organs," Science **335**, 1458–1462 (2012).
20. J. M. Hales, D. J. Hagan, E. W. Van Stryland, K. Schafer, A. Morales, K. D. Belfield, P. Pacher, O. Kwon, E. Zojer, and J.-L. Brédas, "Resonant enhancement of two-photon absorption in substituted fluorene molecules," J. Chem. Phys. **121**, 3152–3160 (2004).
21. J. R. Lakowicz, I. Gryczynski, H. Malak, and Z. Gryczynski, "Two-Color Two-Photon Excitation of Fluorescence," Photochem. Photobiol. **64**, 632–635 (1996).
22. A. Rapaport, F. Szipöcs, and M. Bass, "Dependence of two-photon-absorption-excited fluorescence on the angle between the linear polarizations of two intersecting beams," Appl. Phys. Lett. **82**, 4642–4644 (2003).
23. P. Mahou, M. Zimmerley, K. Loulier, K. S. Matho, G. Labroille, X. Morin, W. Supatto, J. Livet, D. Débarre, and E. Beaurepaire, "Multicolor two-photon tissue imaging by wavelength mixing," Nat. Methods **9**, 815–818 (2012).
24. L.-C. Cheng, N. G. Horton, K. Wang, S.-J. Chen, and C. Xu, "Measurements of multiphoton action cross sections for multiphoton microscopy," Biomed. Opt. Express **5**, 3427–3433 (2014).
25. C. Xu and W. W. Webb, "Measurement of two-photon excitation cross sections of molecular fluorophores with data from 690 to 1050 nm," JOSA B **13**, 481–491 (1996).
26. E. W. Miller, J. Y. Lin, E. P. Frady, P. A. Steinbach, W. B. Kristan, and R. Y. Tsien, "Optically monitoring voltage in neurons by photo-induced electron transfer through molecular wires," Proc. Natl. Acad. Sci. **109**, 2114–2119 (2012).
27. S. T. Flock, S. L. Jacques, B. C. Wilson, W. M. Star, and M. J. van Gemert, "Optical properties of Intralipid: a phantom medium for light propagation studies," Lasers Surg. Med. **12**, 510–519 (1992).
28. G. Hong, S. Diao, J. Chang, A. L. Antaris, C. Chen, B. Zhang, S. Zhao, D. N. Atochin, P. L. Huang, K. I. Andreasson, and others, "Through-skull fluorescence imaging of the brain in a new near-infrared window," Nat. Photonics **8**, 723–730 (2014).
29. C. Wang, R. Liu, D. E. Milkie, W. Sun, Z. Tan, A. Kerlin, T.-W. Chen, D. S. Kim, and N. Ji, "Multiplexed aberration measurement for deep tissue imaging in vivo," Nat. Methods **11**, 1037–1040 (2014).
30. N. Ji, D. E. Milkie, and E. Betzig, "Adaptive optics via pupil segmentation for high-resolution imaging in biological tissues," Nat. Methods **7**, 141–147 (2010).
31. N. Ji, T. R. Sato, and E. Betzig, "Characterization and adaptive optical correction of aberrations during in vivo imaging in the mouse cortex," Proc. Natl. Acad. Sci. **109**, 22–27 (2012).
32. K. Wang, D. E. Milkie, A. Saxena, P. Engerer, T. Misgeld, M. E. Bronner, J. Mumm, and E. Betzig, "Rapid adaptive optical recovery of optimal resolution over large volumes," Nat. Methods **11**, 625–628 (2014).

## 1. Introduction

Two-photon microscopy has had an enormous influence on animal studies of brain activity *in vivo* providing a tool for high-resolution imaging in live cortical tissue [1-4]. Yet, the majority of 2-photon imaging studies, as of today, have focused on the top ~500 µm of the cerebral cortex due to limited penetration of light in biological tissues. Cerebral neurons, however, are wired in circuits spanning the entire cortical depth (~ 1 mm in mice) [5], and sampling of activity throughout this depth would be required for reconstruction of circuit dynamics. Therefore, pushing the depth penetration of microscopic imaging technology is at the heart of the BRAIN Initiative's efforts focused on large-scale recording of neuronal activity [6]. Here, we demonstrate that the degree of freedom introduced by non-degenerate 2-photon excitation (ND-2PE) – using two independently controlled pulsed laser sources of different photon energies – may provide a number of advantages over the conventional methods promising deeper penetration with higher efficiency of excitation in a scattering medium.

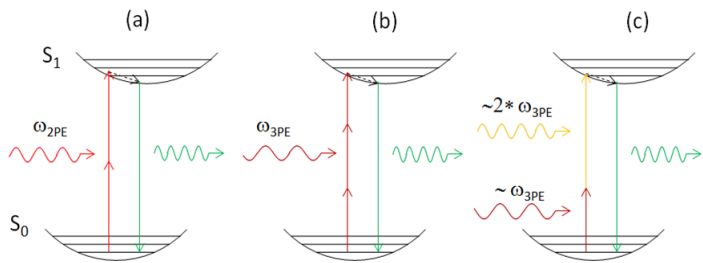

Fig. 1. Schematic energy diagram demonstrating degenerate and non-degenerate multi-photon absorption of a molecule: (a) degenerate 2-photon excitation (D-2PE); (b) degenerate 3-photon excitation (3-PE); and (c) non-degenerate 2-photon excitation (ND-2PE).

The conventional 2-photon microscopy relies on absorption of two equal energy photons (i.e., "degenerate" 2PE (D-2PE), Fig. 1(a)). The light source (usually, a pulsed femtosecond Ti:Sapphire laser) is tuned within the ~740-1000 nm near infrared (NIR) window. For example, 740 nm is used for imaging of nicotinamide adenine dinucleotide (NADH) [7, 8] and 800-1000 nm for imaging of intravascular dextran-conjugated fluorescein isothiocyanate (FITC) [9]. Using even longer excitation wavelengths further pushes the depth limit due to an increase in the photon mean free path [10-12]. A recent study using 1280 nm excitation documented imaging of mouse cortical vasculature *in vivo* down to ~1.6 mm [11] approximately reaching the fundamental depth limit in scattering tissue [13]. This wavelength, however, lies outside the D-2PE spectrum for most of the visible fluorophores, thus limiting the approach to a handful of optical probes with far red emission (e.g., Alexa 680).

Another strategy to increasing imaging depth is the use of higher-order multiphoton excitation. Three-photon imaging relying on absorption of three equal energy photons (i.e., "degenerate" 3-photon excitation (D-3PE), Fig. 1(b)). The short-wavelength infrared (SWIR) spectral excitation window of 1700 nm has been recently demonstrated to penetrate down to ~1.3 mm providing images of hippocampal neurons expressing red fluorescent protein [14]. Higher-order multiphoton excitation, however, suffers from very low absorption efficiency quantified as the absorption cross-section [15]. As such, D-3PE cross-section is orders of magnitude lower than that for D-2PE [16], resulting in low emitted photon count and slowing down image acquisition. Low fluoresence intensity is a significant drawback for real-time imaging of time-resolved biological processes, e.g., neuronal activity. Further, the higher laser power needed to overcome the low D-3PE cross-section may harm tissue resulting from repeated dwelling on the same point as required for imaging of activity over time.

Finally, deeper subcortical imaging can be achieved using more invasive approaches such as microendoscopy[17], by removing the overlaying tissue [18], or with technologies that do not rely on optical focusing but at a price of decreased spatial resolution [19]. However, compromising the health of tissue or spatial resolution might not be an acceptable alternative for most cellular-level neuroscience applications.

To overcome these limitations, we explore ND-2PE – absorption of two photons of different energy (Fig. 1(c)). Non-degenerate excitation of optical species has a long history in the physical chemistry community [20-22] and has been recently employed for simultaneous imaging of multiple fluorophores in the cerebral cortex *in vivo* [23]. For the goal of deep tissue imaging, this mode of excitation holds a promise of attaining higher signal level characteristic to D-2PE with deeper penetration typical for longer wavelengths of SWIR light.

In this study, we investigate ND-2PE fluorescence and show experimentally that its efficiency is comparable with D-2PE. Additionally, we show that for imaging through a scattering medium such as brain tissue, ND-2PE can mitigate the loss due to scattering, and has an effective efficiency substantially higher than that of D-2PE. In the following sections, we describe the experiments in support of the characterization of ND-2PE efficiency and its benefits for operation in a scattering medium. Specifically, our experimental validations start with an investigation of the excitation cross-section of ND-2PE in comparison with that of D-2PE in a transparent medium. The experimental setup is described in Section 2.1, followed by a description of experimental data and their analysis in Sections 2.2-2.4. In Section 2.5, we analyze the advantage of ND-2PE for imaging in a scattering medium (approximating the brain tissue) using the ratio of absorption cross-sections for ND-2PE and D-2PE measured in Sections 2.3 and 2.4 together with the attenuation length for the NIR and SWIR beams from Ref.[14]. Here, we use a model governed by the Beer-Lambert law to predict how far the penetration depth of ND-2PE can be extended to support the minimum detectable fluorescence signal. In Section 2.6, we demonstrate that introducing the SWIR beam in the scattering medium helps to compensate for the scattering loss of the NIR beam, thereby demonstrating that the effective efficiency of ND-2PE is superior to that of D-2PE (i.e., 88% higher efficiency).

## 2. Experimental setup and results

### 2.1 Experiment setup

For our experimental validation, we need to choose one of the beams within the most "transparent" part of SWIR optical window of ~1300-1400 nm or ~1600-1900 nm as has been identified in a recent study [14]. This places the second beam within the ~ 700-1000 nm NIR wavelength range to supply the required photon energy for ND-2PE of visible emission fluorophores (green to red). The NIR beam will experience higher losses due to scattering, limiting the achievable power at the focal spot deep in the tissue. However, under ND-2PE, the emitted signal is proportional to the product of the power in each beam. Therefore, increasing the SWIR power will compensate for the NIR losses. In our experiment, ultrashort pulses were derived from a Ti:Sapphire laser (Coherent Mira-900) tuned to 825 nm, further referred to as the NIR beam) and an optical parametric oscillator (OPO) (1315 nm, further referred to as the SWIR beam).

We choose 1315 nm as the shortest wavelength available from our OPO system while the other wavelength was fixed at 825 nm in order to efficiently operate the OPO system. The NIR beam was separated into two arms by a polarizing beam splitter. One of the arms was launched into the OPO. A delay line (Thorlabs LTS150) with resolution of 2 μm, which is temporally equivalent to 6.6 fs, was used to temporally overlap the NIR and the SWIR laser pulses. Several lenses and mirrors in each optical path were used to spatially overlap the NIR and SWIR laser pulses as indicated in Fig. 2. Then the two laser beams were combined by a dichroic mirror (Thorlabs, DMSP 1000). A lens (Newport, 5712-A-H 40X) was used to focus the NIR and SWIR beams into the sample and collect the emitted fluorescence signal. In order to detect fluorescence, two dichroic mirrors (Semrock, FF678-Di01-25x36 and FF735-Di01-25x36) and a bandpass filter (FF01-530/11) were placed in front of a photomultiplier tube (PMT, Hamamatsu R3896) that was used as a detector.

A computer code (Matlab) was devised to automate the procedure of temporal overlap between the SWIR and NIR pulses by scanning the delay line and logging PMT readings at each position. To achieve the spatial overlap of the two beams, we estimated the position of focal plane for each beam as the following: We inserted a diffuser filter mounted on a translation stage (Newport, M-561D-XYZ) in the light path near the focal plane of the microscope objective and observed the size and motion of the speckle pattern on a target placed behind the filter while translating the filter in the axial direction (along the z axis) and in the transverse (xy) plane. When the filter was translated along the z axis, the speckle size was maximized as it approached the focal plane. When the filter was translated in the xy plane, the moving speckle pattern reversed its direction while passing through the focal plane. Based on these measurements, we estimated that the focal planes of the two beams were displaced by 2 μm along the z axis.

To calculate the beam radius at the focal spot (i.e., beam waist), we used a beam profiler (Thorlabs, BP209-IR). For each of the beams we obtained the beam size at several axial positions for estimation of the divergence angle behind the objective lens [24]. These measurements were then used to back calculate the position of the focal spot. The calculated beam radius at the focal spot (i.e., beam waist) was $w_{NIR} = 1.32$ μm and $w_{SWIR} = 2.135$ μm for the NIR and SWIR beams, respectively.

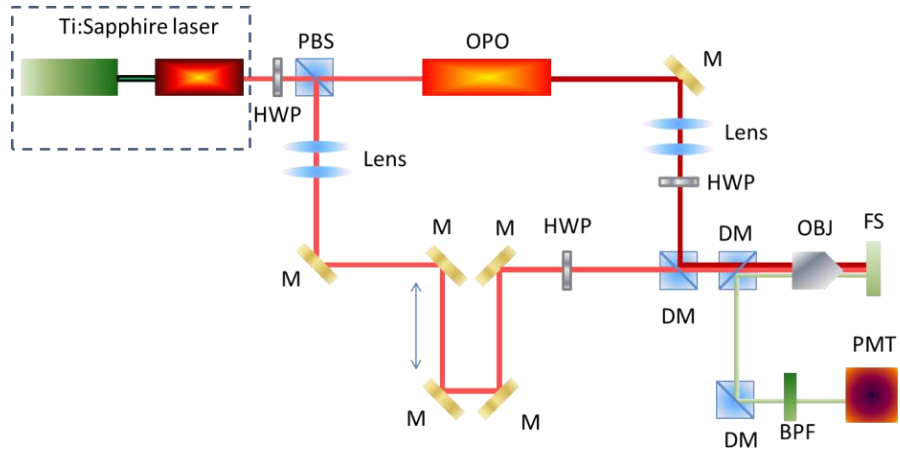

Fig. 2. Experimental setup for demonstration of ND-2PE. PBS, polarizing beam splitter; HWP, half wave plate; DM, dichroic mirror; FS, fluorescent sample; 10XOBJ, microscope objective; BPF, band pass filter; PMT, photomultiplier. M, mirror

*2.2 Fluorescence signal under D-2PE and ND-2PE*

First, we investigated the dependence of the fluorescence signal on the temporal beam alignment (Fig. 3(a)) and power of each beam (Fig. 3(b)-3(c)) in a transparent medium (500 µM fluorescein in saline). We chose fluorescein because it is a common fluorophore with well-characterized multiphoton absorption properties [25] and is the basis for novel voltage sensitive probes [26]. The temporal alignment was optimized by scanning the optical delay line, and the fluorescence signal was detected by the PMT. Irrespective of the delay line position, we observed a fluorescence signal generated by the D-2PE due to the NIR beam alone (Fig. 3(a), "D-2PE"). No fluorescence signal was generated due to the SWIR beam alone because of its insufficient photon energy for D-2PE of fluorescein. The delay line was moved in steps of 2µm using a stepper motor. This step size was chosen considering the pulse width of the NIR and SWIR beams which were around 146 and 185 fs respectively (i.e., 45 and 60 µm in free space, respectively). An increase in the fluorescence signal was detected as the two beams overlapped indicating the additive effect of the ND-2PE occurring due to the simultaneous absorption of NIR and SWIR photons (Fig. 3(a), "D-2PE + ND-2PE"). This additive property indicates that, under our experimental conditions, depletion of NIR beam due to its absorption by the fluorophore in the D-2PE regime was insignificant. The ND-2PE at $\lambda_1=825$ nm and $\lambda_2=1315$ nm is equivalent in photon energy to D-2PE with $\lambda_3=2/(1/\lambda_1+1/\lambda_2)=1013$ nm.

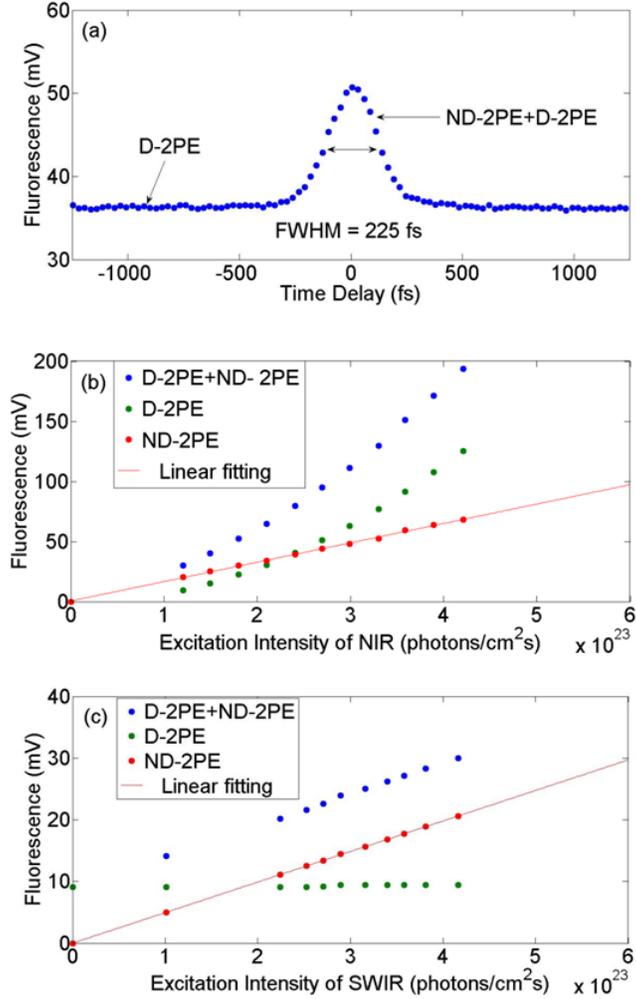

Fig. 3. Experimental demonstration of ND-2PE. (a) dependence of the fluorescence signal on temporal alignment; (b) fluoresence intensity dependence on NIR power (SWIR intensity is fixed at $4.16\times10^{23}$ photons/cm$^2$s); (c) fluorescence intensity dependence on SWIR power (NIR intensity is fixed at $1.18\times10^{23}$ photons/cm$^2$s).

*2.3 Power dependence of ND-2PE and D-2PE on NIR and SWIR beam*

Next, we examined the dependence of the fluorescence signal on the power of each beam (Fig. 3(b)-(c)). In the absence of proper temporal overlap, the signal scaled quadratically with the NIR power with no dependence on the SWIR power (green curves in Fig. 3(b) and Fig.3(c), respectively). This clearly indicates D-2PE. With the two beams aligned, the excited fluorescence signal increased with an increase in power of each beam while keeping the power of the other one constant (blue curves in Fig. 3(b) and Fig. 3(c)). Subtracting the contribution of D-2PE we obtained a dependence of the excited fluorescence on the power of each beam (red curves in Fig. 3(b) and Fig. 3(c)). As expected, the signal increased linearly in both experiments clearly indicating ND-2PE (that is proportional to the power in each beam). The measured power dependences in Fig. 3(c) do not start from zero because of the fluorescence due to D-2PE at the fixed NIR power. 3PE was not observed due to insufficient SWIR power emphasizing low efficiency of 3PE compared with ND-2PE [24].

*2.4 Cross section of ND-2PE*

To quantify the absorption efficiency, we computed the ratio of absorption cross-sections for D-2PE ($\sigma_D$) and ND-2PE ($\sigma_{ND}$). To this end, we started with the expression of fluorescence while assuming nonuniform intensity within each beam:

$$F_D(z) = K \iint \sigma_D \cdot I_{NIR}(t,r,z) \cdot I_{NIR}(t,r,z) dr dt , \qquad (1a)$$

$$F_{ND}(z) = K \iint \sigma_{ND} \cdot I_{NIR}(t,r,z) \cdot I_{SWIR}(t,r,z) dr dt . \qquad (1b)$$

where $F_D$ and $F_{ND}$ are the detected fluorescence signal under D-2PE and ND-2PE, respectively; K is a product of the quantum yield of the fluorophore, geometry of the imaging system and the fluorophore concentration and is independent of the excitation regime; $I_{NIR}(t,r,z)$ and $I_{SWIR}(t,r,z)$ are the spatiotemporal intensity distributions assuming Gaussian beams in space and Gaussian pulse envelope. After simplification, the ratio of these two Eqs yields

$$\frac{F_{ND}}{F_D} = \frac{\sigma_{ND} \cdot I_{NIR} \cdot I_{SWIR} \cdot \alpha}{\sigma_D \cdot I_{NIR} \cdot I_{NIR}} . \qquad (2)$$

where $I_{NIR}$ and $I_{SWIR}$ are the average photon flux for the SWIR and NIR beams and α is a correction factor that accounts for the partial overlap in the SWIR and NIR beam due to differences in the pulse width and beam waist. If we adjust the beam power, $P_{NIR}$ and $P_{SWIR}$, such that $F_D = F_{ND}$, using Eq (2) we obtain

$$\frac{F_{ND}}{F_D} = \frac{\sigma_{ND} \cdot I_{SWIR} \cdot \alpha}{\sigma_D \cdot I_{NIR}} . \qquad (3)$$

The average photon flux, $I_{NIR}$ and $I_{SWIR}$, for each beam was approximated given the power, $P_{NIR}$ and $P_{SWIR}$, photon energy, $\hbar\omega_{NIR}$ and $\hbar\omega_{SWIR}$, and the calculated beam radius $w_{NIR}$ and $w_{SWIR}$ as

$$I_{NIR} = \frac{P_{NIR}}{\hbar\omega_{NIR}} \cdot \frac{1}{\pi w_{NIR}^2} , \qquad (4)$$

$$I_{SWIR} = \frac{P_{SWIR}}{\hbar\omega_{SWIR}} \cdot \frac{1}{\pi w_{SWIR}^2} . \qquad (5)$$

This calculation yielded $\alpha = 1.10$ and $\sigma_{ND}/\sigma_D = 0.51 \pm 0.1$. This value for the $\sigma_{ND}/\sigma_D$ ratio is most likely explained by the off-resonance ND-2PE excitation (equivalent to D-2PE at 1013 nm). These results confirm that the power requirement for the NIR beam, which would experience higher scattering in biological tissue, can be relaxed at the expense of increasing the SWIR power.

*2.5 Analysis of the penetration depth with ND-2PE in scattering medium*

The results of the previous sections show that the excitation cross-section for ND-2PE in the transparent medium is about half of that for D-2PE. However, we anticipate that effectively the efficiency of ND-2PE will be significantly higher compared to that of D-2PE in the scattering medium due to the difference in attenuation length for the NIR and SWIR beams. Next, we performed a theoretical calculation to demonstrate the benefits of ND-2PE for imaging in a scattering medium. For this calculation, we approximated the attenuation length of our excitation beams using previously published values for NIR and SWIR at comparable wavelengths (130 μm and 285 μm, for NIR=775nm and SWIR=1280nm, respectively)[14]. We modeled the beam attenuation with the Beer-Lambert law using the previously mentioned attenuation coefficients. In order to reveal the effect of ND-2PE, for simplicity, the photon intensity of the NIR and SWIR beams were set to be equal resulting in equal total excitation power for the D-2PE and ND-2PE before entering the scattering medium. The fluorescence signal in the simulation was normalized to the fluorescence signal of D-2PE at the entrance to the scattering medium (at the depth of 0 μm). The calculated result is shown in Fig. 4, where the blue and green curves represent the fluorescence intensity resulting from D-2PE and ND-2PE, respectively. Within the attenuation length of the NIR beam, the lower fluorescence signal of ND-2PE is a result of its lower excitation cross-section (as measured in Sections 2.2-2.4). Past this point, the advantage of ND-2PE becomes obvious. Since the fluorescence intensity depends quadrically on the intensity of the NIR beam which experiences higher scattering, the ND-2PE signal begins to exceed that of D-2PE. This is because the SWIR beam is attenuated less during propagation in the scattering medium. At a depth of 700 μm, where the lowest signal could be detected experimentally in Ref. [14], the signal drops by ~47dB. For the ND-2PE case, however, the same ~47 dB loss is calculated to occur at ~900 μm, a ~200 μm enhancement of the penetration depth. In addtion, we can think of this effect in terms of efficiency. At a depth of 700 μm, the fluorescence of D-2PE drops by ~47 dB while the fluorescence of ND-2PE drops by only ~37 dB. The 10-dB difference is due to ND-2PE being relatively unaffected by the loss of the NIR power in the scattering medium. This aspect is the core of the following experiment in Section 2.6.

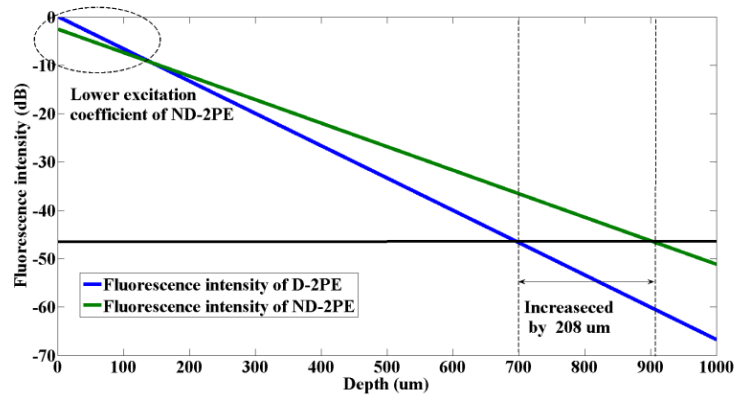

Fig, 4 Simulation of the fluorescence intensity under D-2PE (blue) and ND-2PE (green) as a function of depth in the scattering medium.

*2.6 Efficient ND-2PE in scattering medium*

Next, we conducted a separate experiment comparing the fluorescence signal generated using ND-2PE and D-2PE in the transparent and scattering media. The excitation intensity of the NIR and SWIR beams was increased beyond the range of the previous experiment ($1.66 \times 10^{24}$ and $3.16 \times 10^{24}$ photons/cm$^2$s for the NIR and SWIR, respectively) to overcome losses in the scattering medium. To emulate the scattering properties of biological tissue we used a solution of 1 % intralipid in water (by volume), which is a standard phantom for studies of light penetration in biological tissues [27, 28]. The sample (500 μM fluorescein) was submerged 500 μm below the surface of the scattering medium. Similar to the experiment in the transparent medium, we moved the delay line and measured the fluorescence signal as a function of the delay line position (Fig. 5 (a)). The resultant curve was then normalized by the fluorescent signal ($F_D$) due to D-2PE when the beams were temporally misaligned (Fig. 5(b)) resulting in the ratio $F_{ND}/F_D$ (see Eq. (2)) in the transparent and scattering media (blue and green curves, respectively). An $F_{ND}/F_D > 1$ indicates a fluorescence intensity enhancements resulting from ND-2PE. Since the SWIR experiences less scattering in the intralipid solution, the normalized fluorescence signals under ND-2PE (green curve) was effectively enhanced by 88% compared to that in the transparent medium (blue curve). This result provides a practical demonstration that loss of the NIR power in the scattering medium can be efficiently compensated for by increasing the SWIR power.

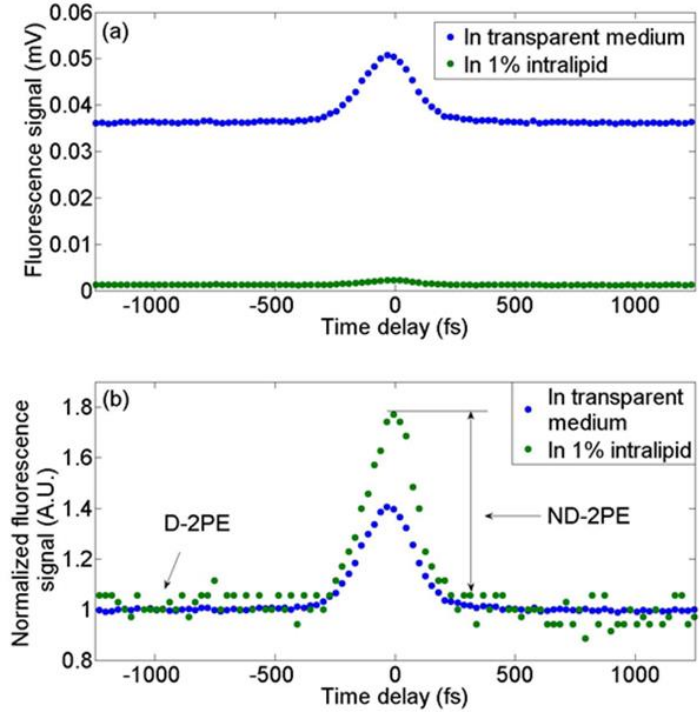

Fig. 5. (a)Fluorescence intensity as a function of the temporal alignment between NIR and SWIR beams in the transparent medium (blue curve) and in 1% intralipid (green curve). (b) The same as (a) where each curve has been normalized by the respective D-2PE signal.

## 3. Discussion and conclusions

In this study we demonstrated the advantage of ND-2PE providing higher efficiency of excitation in the scattering medium. Increasing the photon density of the SWIR beam helps to compensate for the scattered loss of the NIR beam. However, this effect is limited by the maximum laser output and the damage threshold of biological tissues. Further studies will be required to determine the optimal combination of powers for achiving the maximum fluorescence signal while avoiding tissue damage. Regardless of the damage threshold, further improvement in the excitation efficiency can be achieved by implementation of Adaptive Optics (AO) to correct the phase distortions experienced by the NIR beam [29-31]. Specifically, the SWIR beam, which can be focused deep inside the tissue, can be used as a reference point ("guiding star") allowing us to adjust the phase of the NIR beam to achieve the maximum spatial overlap of their focus spots. An NIR beam has been recently introduced to correct distortions of the focal spot in conventional (degenerate) 2-photon microscopy [32]. The same correction procedure will also compensate for chromatic aberrations of the objective allowing implementation of standard objectives used in degenerate 2-photon microscopy for non-degenerate excitation. Further, the SWIR and NIR beams can be strategically displaced to ensure they overlap only at the focal volume avoiding unwanted "out-of-focus" excitation on the brain surface [13].

The current study was limited to a fixed combination of the SWIR and NIR photon energies producing the effect equivalent to excitation at 1013 nm. In the future, efficiency of excitation (i.e., excitation cross-section) can be increased through a systematic search for the most efficient combination [20]. The optimal combination as well as the degree of improvement will vary among fluorophores depending on their chemical structure.

To summarize, ND-2PE may provide a superior alternative to 3-photon microscopy by circumventing the issue of low 3PE cross-section while taking advantage of low scattering SWIR illumination and AO. With the 3PE regime as a reference, we can expect orders of magnitude improvement in the excitation cross-section. Compared to the degenerate 2-photon microscopy, this technology will provide the advantage of larger penetration depth similar to that under 3PE. Further increases in excitation efficiency due to the phenomenon of resonant enhancement will apply to some fluorophores. In the long run, we envision a parallel development of non-degenerate 2-photon microscopy instrumentation and molecular engineering of organic molecules and fluorescent proteins tailored for ND-2PE.

Acknowledgments.

This research was supported by UCSD Center for Brain Activity Mapping (CBAM) Seed grant, NSF, ONR, NIH (NS057198, EB00790 and S10RR029050) and Cymer Corporation.